\documentclass[12pt, a4paper]{article}
\usepackage{amssymb, amsthm, amsmath, setspace}
\usepackage{xcolor, tikz, mathpartir, mathtools}
\usepackage{biblatex, hyperref, cleveref}

\hypersetup{
  colorlinks,
  linkcolor={red!50!cyan},
  citecolor={purple!50!brown},
  urlcolor={blue!80!brown}
}

\newcommand{\Prop}{\textsf{Prop}}
\newcommand{\Tp}{\textrm{Tp}}
\newcommand{\Tm}{\textrm{Tm}}
\newcommand{\PropPsh}{\textrm{Prop}}
\DeclareMathOperator{\El}{El}
\DeclarePairedDelimiter\name{\textnormal{“}}{\textnormal{”}}
\newcommand{\strong}[1]{\textbf{#1}}

\title{A Coherence Construction for the Propositional Universe}
\author{Huang Xu}
\date{}

\begin{document}
\maketitle
\begin{abstract}
We record a particularly simple construction
on top of Lumsdaine's local universes~\cite{Lumsdaine:LocalUniverse}
that allows for a Coquand-style universe of propositions
with propositional extensionality to be interpreted
in a category with subobject classifiers.
\end{abstract}

\section{Introduction}

The models of type theory usually require equations
closely following that in the syntax.
This makes it easy to establish the relation of syntax and semantics.
However, with dependent types,
equalities of terms inevitably require equalities on types,
which are intuitively interpreted as objects in a category,
and hence violate the principle of equivalence.
In particular, a lot of naturally occurring mathematical objects
will not satisfy the equalities on the nose,
merely having isomorphisms between the objects.
This means we cannot use these objects directly
to construct models of type theory.

To bridge the gap, numerous constructions can be used
to turn a model where the equalities are satisfied up to isomorphism
to one where the equalities strictly hold.
These include the Hofmann–Streicher construction~\cite{Hofmann:LCCC}
and Lumsdaine's local universes model~\cite{Lumsdaine:LocalUniverse}.

In this note, we write down a construction that
allows us to interpret a Coquand-style universe \(\Prop\) of propositions
as a subobject classifier \(\Omega\).
This can be applied, for example, to interpret extensional type theory
with \(\Prop\) in a topos.

\section{Preliminaries}

A \strong{Coquand universe} is a type \(\mathcal{U}\)
equipped with two operations \(\El(-)\) and \(\name{-}\).
The elements of \(\mathcal{U}\) is called the \strong{codes} of types.
\(\El(\alpha)\) turns a code \(\alpha\) to its corresponding type,
and \(\name{A}\) turns a type into its code,
provided that there exists such a code in \(\mathcal{U}\).
\begin{mathpar}
\inferrule{
  \Gamma \vdash \alpha : \mathcal{U}
}{
  \Gamma \vdash \El(\alpha)~\textrm{type}
} \and
\inferrule{
  \Gamma \vdash A~\textrm{type} \\
  \text{(...)}
}{
  \Gamma \vdash \name{A} : \mathcal{U}
}
\end{mathpar}
where the ellipsis represent additional side conditions
that guarantee \(A\) has a code in \(\mathcal{U}\),
which depends on the specifics of \(\mathcal{U}\).
Moreover, these two operations are mutual inverses.
\[
\Gamma \vdash \El(\name{A}) = A~\textrm{type} \qquad
\Gamma \vdash \name{\El(\alpha)} = \alpha : \mathcal{U}
\]
We then introduce rules that relate the codes to the types as desired.
For example, a type theory with binary products can
introduce a constructor
\(\textsf{Pair} : \mathcal{U} \to \mathcal{U} \to \mathcal{U}\),
such that \(\El(\textsf{Pair}(\alpha, \beta)) = \El(\alpha) \times \El(\beta)\).

Propositions in type theory have many formulations.
Here, we use the term as referring to a type
with at most one element, in a broad sense.
In particular, this means any two elements of the type are equal.
This can be interpreted as a judgemental rule
\[\inferrule{\Gamma \vdash x : A \\ \Gamma \vdash y : A}{\Gamma \vdash x = y : A}.\]
It can also be interpreted as a provable statement
\(\forall(x\,y : A), x = y\).
In the case we are interested in, i.e. extensional type theory,
they have no difference.
A universe of propositions is a universe whose types are all propositions.
This intuitively corresponds to the subobject classifier in a category.

Propositional extensionality states that
two logically equivalent propositions are actually equal.
In other words, if \(p \to q\) and \(q \to p\), then \(p = q\).
This can be seen as a special case of the univalence principle,
but it is also very useful in non-univalent formal systems.

We use natural models~\cite{Awodey:NatMod}, or categories with families,
as our working notion of models of type theory.
In short, this is a category equipped with two presheaves \(\Tp, \Tm\)
and a natural transformation \(\tau : \Tm \to \Tp\).
The category represents the contexts and substitutions between them.
For each object \(\Gamma\), the set \(\Tp(\Gamma)\) represents
the set of types in the context \(\Gamma\).
The set \(\Tm(\Gamma)\) represents
the well-typed terms up to judgemental equality,
and \(\tau : \Tm(\Gamma) \to \Tp(\Gamma)\)
assigns the corresponding types to the terms.
The reader can refer to Awodey~\cite{Awodey:NatMod}
for a more detailed introduction.

\section{Coherence}

A Coquand universe can be added to
the definition of natural models with relative ease.
A type \(\mathcal{U}\) corresponds to a point
\(U : 1 \to \Tp\) in the presheaf of types.
The pullback of \(U\) with \(\tau\) is then
a presheaf \(\tilde{U}\) representing elements of \(\mathcal{U}\).
The operation \(\El\) corresponds to
a natural transformation \(\tilde{U} \to \Tp\).
On the other hand, \(\name{-}\) corresponds to
a natural transformation in the reverse direction.
Of course, with the side condition on \(\name{-}\)
as mentioned earlier, we just need to restrict the domain
to some subpresheaf of \(\Tp\) as appropriate.

In our case, we have a subpresheaf \(\PropPsh \hookrightarrow \Tp\)
of subsingleton types.
And the two natural transformations constitutes
a natural isomorphism \(\tilde{U} \cong \PropPsh\).
In the language of diagrams, a family of types
\(X \to \Tp\) is a family of subsingletons if and only if
the pullback of \(\tau : \Tm \to \Tp\) is a monomorphism
\(\tilde{X} \to X\).
In Lumsdaine's model, a subsingleton type in the context \(\Gamma\)
is a diagram \(\Gamma \to V \leftarrow V_*\) in the base category
whose pullback is a monomorphism.

Given a subobject classifier \(\Omega\),
it's tempting to try to interpret the propositional universe
as the diagram \(\Gamma \to 1 \leftarrow \Omega\).
However, this does not work, because its elements
corresponds to subobjects of \(\Gamma\),
which is generally not in bijection with subsingleton types.
We only obtain a \emph{retract} \(\tilde{U} \hookrightarrow \PropPsh\) in this way.
More concretely, \(\El\) maps a code (which is a subobject \(P \hookrightarrow \Gamma\))
to the diagram \(\Gamma \mathrel{\smash{\xrightarrow{\operatorname{id}}}} \Gamma \hookleftarrow P\),
and \(\name{-}\) maps a diagram to a chosen pullback map.

To remedy this, we quotient our \(\Tp\) by a carefully chosen equivalence.
Given a type \(A \in \Tp(\Gamma)\),
we have a presheaf \(A^\circ\) on the slice category over \(\Gamma\),
given by sending \(\sigma : \Delta \to \Gamma\) to
the set of elements of \(A[\sigma] \in \Tp(\Delta)\).
Two types are judgementally isomorphic if
their corresponding presheaves are naturally isomorphic.
We consider two types equal iff either
(a) they are the same, or
(b) their corresponding presheaves are subsingletons,
and they are judgementally isomorphic.

By construction,
our equivalence relation automatically respects
the functorial action of substitutions.
Suppose \(A, B \in \Tp(\Gamma)\) are judgementally isomorphic,
i.e. \(A^\circ \cong B^\circ\),
then given a substitution \(\sigma : \Delta \to \Gamma\),
\(A[\sigma]^\circ\) sends \(\delta : \Xi \to \Delta\) to
the elements of \(A[\sigma][\delta] = A[\sigma \circ \delta]\).
Hence the natural isomorphism \(A^\circ \cong B^\circ\)
induces a natural isomorphism \(A[\sigma]^\circ \cong B[\sigma]^\circ\).

Therefore we obtain a quotient presheaf \(\overline{\Tp}\).
Most importantly, we can construct a corresponding \(\overline{\Tm}\),
which will act as the new presheaves of terms and types.
This is because subsingletons have only one automorphism.
The elements of \(\Tp\) that are identified together
have at most one element in \(\Tm\) mapping to them,
so there we can simply identify those elements in \(\Tm\) too.

After the quotient,
isomorphic propositions in \(\PropPsh\) will become equal,
and therefore \(\PropPsh \cong \tilde{U}\) will hold.
This also enables propositional extensionality.

\section{Discussion}

We have provided a coherence construction
that enables us to interpret a proposition universe
à la Coquand as the subobject classifier in a category.
As we can see from the construction,
the fact that no non-trivial automorphisms exist
for propositions is essential for the quotient to work.

In the general case,
a more complicated construction is required to kill off the automorphisms.
This can be seen for instance in the construction of
a simplicial model for homotopy type theory~\cite{Kapulkin:Simplicial}:
in section 2.1, the authors equipped Kan complexes with well-orderings,
a structure well-known to be rigid.
The construction in this article may be more suitable
for simpler use cases where other approaches are an overkill.

\printbibliography

\end{document}